\magnification= \magstep 1
\openup 1 \jot
\vsize=23.5true cm
\hsize=16true cm
\nopagenumbers
\topskip=1truecm
\headline={\tenrm\hfill\folio\hfill}
\raggedbottom
\abovedisplayskip=3mm
\belowdisplayskip=3mm
\abovedisplayshortskip=0mm
\belowdisplayshortskip=2mm
\normalbaselineskip=12pt
\normalbaselines

\centerline {\bf Self-gravitating magnetic monopoles,}
\centerline {\bf global monopoles and black holes}
\vskip 2cm
\centerline {G W Gibbons}
\vskip 1.5cm
\centerline {D.A.M.T.P.}
\centerline {University of Cambridge}
\centerline {Silver Street}
\centerline {Cambridge CB3 9EW}
\vskip 2.5cm

{\bf Introduction}. The properties of the simplest magnetic monopoles in
Yang-Mills-Higgs theory in
flat spacetime are by now quite well understood.  Moreover the fact that almost
any grand unified theory will admit monopole solutions has had a profound
impact on modern cosmology since it provides a very strong argument in favour
of inflation as a possible solution of the monopole problem.  It is therefore
rather surprising that comparatively little is known about the gravitational
properties of magnetic monopoles and the relation between monopoles and black
holes.  In the past couple of years or so this situation has begun to change
and it is my purpose in these lectures to review the current situation and make
some comments on it. I have included a fairly complete list of references, not
all of which are referred to  explicitly in the text, for the convenience of
those wishing to follow up these topics.
\vskip 2.0cm
\leftline {\bf {Contents}}
\vskip .5 cm
1.  Monopoles in Flat Spacetime

2.  Static Solutions of Einstein's Equations without Horizons

3.  Bogomolnyi Bounds for Einstein-Yang-Mills-Higgs initial data

4.  Static solutions of the Einstein-Yang-Mills Equations with Horizons

5.  Solutions of the Einstein-Yang-Mills-Higgs Equations with Horizons

6.  Global Monopoles and Black Holes

7.  Black Hole Monopole Pair Production

\vfill
\eject

{\bf 1.  Monopoles in Flat Spacetime}.  The virial theorem tells us that any
finite
energy time-independent solution of the equations of motion must satisfy:
$$
\int _ {I \negthinspace R ^ 3} T _ {ik} d ^ 3 x = 0
\eqno (1.1)
$$
where $T _ {ik}$ are the spatial components of the energy-momentum tensor.
Equation (1.1) follows from the conservation equation $T _ {ij,j} = 0$ and the
obvious consequence:
$$
(T_{ij} x_k), j = T _ {ik}
$$
using the divergence theory and discarding the boundary term.  The physical
meaning of (1.1) is that the total stresses in an extended object must balance.
In particular the components of $T_{ik}$ cannot have a fixed sign - for example
the spatial trace $\sum _ i T _ {ii}$ equals the sum of the \lq \lq principle
pressures\rq \rq so there must be regions where the matter is in tension and regions
where it is in compression.  For a pure Higgs field (assumed through out these
lectures to lie in the
adjoint representation of $SU(2)$)
$$
T _{ij} = + D _ i  {\Phi}. D _ j \Phi - {1 \over 2} \delta _ {ij} (D
_ k \Phi. D _ k \Phi) - {1 \over 2 } \delta _ {ij} \ W (\Phi)
\eqno (1.2)
$$
and (1.1) cannot possibly be satisfied as long as $W(\Phi) \geq 0$.  The
potential term gives an isotropic negative pressure.  The gradient term gives a
positive pressure along the gradient direct and equal tensions in the
orthogonal directions.  The sum of principal pressures is thus negative.  In
fact this result holds for an arbitrary harmonic map with non-negative
potential.

By contrast for a Yang-Mills field
$$
T _ {ij} = -  {B} _i .  {B} _ j + {1 \over 2}  {B}
_ k .   {B} _ k \delta _ {ij}
\eqno (1.3)
$$
where
$$
  {B} _ i = {1 \over 2} \epsilon _ {ijk} F _ {jk}
\eqno (1.4)
$$
is the magnetic field strength.  As Faraday taught Maxwell there is now a {\bf
{tension}} along the direction of $  {B} _ i$ and an equal pressure
orthogonal to the field lines.  The sum of the principal pressures is now
positive.  Thus for pure Yang-Mills there can be no static solution either.

However as 't Hooft and Polyakov showed there {\bf {is}} a solution of the
combined Yang-Mills Higgs system which does satisfy (1.1) and which is stable.
Infact it minimizes the \lq \lq total energy\rq \rq :
$$
\int T_{00} d ^ 3 x =\int d ^ 3 x [{1 \over 2} B ^ 2 + {1 \over 2} (D \Phi) ^ 2
+ W (\Phi) ]
\eqno 1.5
$$
where $D$ is of course now a gauge covariant derivative.  Moreover if $\Phi$
transforms by the adjoint representation of the gauge group $SU(2)$ the total
energy is bounded below by
$$
{{4 \pi \eta} \over e}  = g \eta
\eqno (1.6)
$$
where $e$ is the gauge coupling constant, $g = 4 \pi/e$ the magnetic charge of
the monopole and $|\Phi| \rightarrow \eta$ at infinity.  This Bogomolnyi bound
can only be attained if $W(\Phi)$ vanishes identically (the Prasad Sommerfeld
limit) and the Bogomolnyi equations:
$$
\pm B_i =D _ i \Phi
\eqno (1.7)
$$
hold, moreover from (1.7), it follows that the stresses vanish pointwise i.e.
$T _ {ij} = 0$.

The Prasad Sommerfeld limit and the associated Bogomolnyi equations are of
great mathematical interest and have many geometrical consequences.  They are
also related to supersymmetry:  the system with $W (\Phi) = 0$ admits an $N =
4$ supersymmetric extension.  It has, as a consequence, received a great deal
of attention.  Of greater physical relevance is the case when
$$
W ( \Phi) = \lambda (\Phi ^ 2 - \eta ^ 2) ^ 2
\eqno (1.8)
$$
and the total energy is given by
$$
{4 \pi \eta \over e} f (\lambda / e ^ 2)
\eqno (1.9)
$$
where $f (\lambda / e ^ 2)$ is a dimensionless function of the dimensionless
ratio $\lambda /e ^ 2$ with $f(0) = 1$.

In addition to the finite energy solutions there is a static solution of the
pure Higgs field equations (with $\Phi$ again a triplet of $SU(2)$) $\Phi$
satisfies a Hedgehog Ansatz:
$$
\Phi ^ a = (x ^ a / r) F (r)
\eqno (1.10)
$$
with $F (0) = 0$ and $F (\infty ) = \eta$.  Since
$$
T _ {00} \rightarrow {\eta ^ 2 \over r ^ 2}
\eqno (1.11)
$$
this solution has infinite energy.  It is called a {\bf {global monopole}}.

It is a non-singular solution of the Higgs equations of motion.  It is however
well approximated by a singular solution of the non-linear $\sigma$-model
obtained by enforcing the constraint that
$$
|\Phi|^2 = \eta ^ 2
\eqno (1.12)
$$
everywhere.  Such singular solutions arise in the theory of nematic liquid
crystals.  They are known to be unstable in that a lower energy configuration
is available with the energy concentrated along lines (\lq \lq strings \rq \rq) with energy
per unit length equal to $4 \pi \eta ^ 2$.  Global monopoles have recently been
considered in connection with \lq \lq cosmic textures\rq \rq .  I will discuss them further
in a later section.  Let us first see to what extent these basic facts are
modified when we consider self-gravitating monopoles.
\vfill
\eject
{\bf 2.  Static Solutions of Einstein's Equations Without Horizons}

Globally static metrics (i.e. time independent, time reversal invariant and
without event horizons) may be cast in the form:
$$
ds ^ 2 = -V^2 (x) dt ^ 2 + g _ {ij} (x) dx ^ i dx ^ j
\eqno (2.1)
$$

The field equations are then:
$$
\nabla _ g ^ 2 V = 4 \pi GV (T_ {\hat 0 \hat 0} + \sum T _ {\hat j \hat j})
\eqno (2.2)
$$
$$
R _ {ij} [g] = V ^ {-1} \nabla _ i \nabla _ j V + 4 \pi G g _ {ij}
(T _ {\hat 0
\hat 0}-  \sum   T _ {\hat j \hat j}) + 8 \pi G T _ {ij}
\eqno (2.3)
$$
where $\nabla _ g ^ 2$ is the Laplacian of $g _ {ij}$ and $\nabla _ i$ its
covariant derivative.  $T _ {\hat 0 \hat 0}$ and $T _ {\hat j \hat k}$ are the
components of the energy momentum tensor in an orthonormal frame with $e _ 0 =
V ^ {-1} {\partial \over \partial t}$.  If the
metric is asymptotically flat then
$$V \sim 1 - 2Gm/r+ O ({1 \over r^2})
\ {\rm {and}} \ g _ {ij} \sim (1 + {2Gm \over r }) \delta _ {ij} + 0 ({1 \over r ^
2})
\eqno (2.4)
$$
where $m$ is the A.D.M. mass of the spacetime.  From (2.2) we have
$$
m = \int _ \Sigma \ V (T _ {\hat 0 \hat 0} + \sum  T _ {\hat i \hat i} )
\sqrt {g} d ^ 3 x
\eqno (2.5)
$$
where $\Sigma$ is a surface of constant time (assumed complete).

For some purposes it is convenient to rescale the 3-metric $g _ {ij}$ and
re-write (2.1) as
$$
ds ^ 2 = - e ^ {2U} dt ^ 2 + e ^ {-2U} \gamma _ {ij} dx ^ i dx ^ j
\eqno (2.6)
$$
The quantity $U$ may be called the Newtonian potential.  The field equations
now become:
$$
\nabla ^ 2 _ \gamma U = 4 \pi G e ^ {-2U} (T _ {\hat 0 \hat 0} + \sum _ {\hat i} T
_ {\hat i \hat i} )
\eqno (2.7)
$$
and
$$
\tilde R _ {ij} [\gamma] = 2 \partial _ i U \partial _ j U +
+ 8 \pi G (T _ {ij} -  \gamma _ {ij} e ^ {-2U} \sum  T _ {\hat
i \hat i})
\eqno (2.8)
$$
where $\tilde R _ {ij}$ is the Ricci tensor of $\gamma _ {ij}$.  Now note:

(1)  from (2.4) it follows that $\gamma _ {ij}$ is a complete asymptotically
flat 3-metric with zero ADM mass

(2)  from (2.8) the Ricci scalar $\tilde R$ of $\gamma _ {ij}$ is given by
$$
\tilde R = 2 \gamma ^ {ij} \partial _ i U \partial _ j U - 16 \pi G e^{-2U}
(\sum
 T_{\hat i \hat i} )
\eqno (2.9)
$$

Using the positive mass theorem we now deduce the following

{\bf Theorem 1 {There are no globally static asymptotically flat solutions of
Einstein's
with $\sum _ i T _ {\hat i \hat i} \leq 0$.}}  In other words since gravity is
attractive we need some pressure to resist collapse inwards.  Note that to
prove theorem 1 we do {\bf {not}} need to assume that, the matter has positive
energy.

If $T _ {\alpha \beta} = 0$ theorem 1 is just Lichnerowicz's theorem
If the matter is a scalar field however we obtain a new result:

{ \bf Cor 1. { There are no globally static asymptotically solutions of Einstein's
equations with a minimally coupled scalar field source with a non-negative
potential.}}

It is interesting to note that the solutions recently derived by Vilenkin and
Bariola giving the gravitational fields of global monopoles escape cor. 1 by
virtue of {\bf {not}} being asymptotically flat, as I shall describe later.  It
is also important to point out that there {\bf {do}} exist solutions in which a
complex scalar field varies harmonically with time in such a way that $T _ {\mu
\nu}$ and the metric $g _ {\mu \nu}$ are static.  Such scalar fields
are said to consist of $Q$-matter.

On the other hand for pure Einstein-Yang-Mills we cannot deduce from Theorem 1
that there are no static solutions without horizons, since $\sum T _ {\hat i
\hat i} \geq 0$.  In fact for pure Einstein-Maxwell theory there are in fact
no static solutions without horizons (for a proof see Breitenlohner, Gibbons
and Maison (1988)).  It came as a
surprise therefore when Bartnik and McKinnon (1988) announced the existence
of an integer's worth of static spherically symmetric solutions.  Their metric
ansatz was
$$
ds^2=-e^{2U(r)}dt^2+ {{dr^2} \over {1-{2Gm(r) \over r} }} + r^2 (d \theta
^ 2 + \sin ^ 2 \theta  d \phi ^ 2)
\eqno (2.10)
$$
with
$$
eA = a(r)\tau_3 dt+ w(r) (\tau _ 1 d \theta + \tau _ 2 \sin \theta d \phi) +
\tau _ 3 \cos
\theta d \phi
\eqno (2.11)
$$
where $\tau _ 1, \ \tau _ 2 \ \tau  _ 3$ is the usual basis for the Lie algebra of
$SU(2)$.  Equation (2.11) gives spherically symmetric $SU(2)$ connection over
$S^2$ if we set $r =$ constant, $t$ = constant.  Purely magnetic solutions
have $a(r) = 0$ and $eF = w ^{\prime}dr \wedge d \theta \tau _ 1 + w^{ \prime}
 \sin \theta d
r\wedge d \phi \tau _ 2 - (1 - w ^ 2) \sin \theta d \theta \wedge
d \phi \tau _ 3$.

According to Bartnik and McKinnon the assumption of suitable asymptotics and
of finite energy implies that the electric potential $a (r)$ must vanish.
There results a system of radial ordinary differential equations for $U(r), \
w(r), $ and $m(r)$.  If $^{\prime}$ denotes differentiation with respect to $r$ we
have:
$$
{{m^{\prime}} \over {4 \pi G} }= (1 - {2Gm \over r} )(w^{\prime}) ^ 2 + {1 \over 2} {1 \over r ^ 2}
(1 - w ^2)
$$
$$
r ^ 2 (1 - {2Gm \over r}) w ^{\prime \prime}+ (2Gm -
{(1 - w^2) \over r ^ 2} ^ 2)
w ^{\prime}+
(1 - w ^ 2 ) w = 0
$$
and $$
U = {1 \over 2} \ln (1 - {2Gm \over r}) - {4 \pi G }\int ^ \infty _ r {1 \over r} (w
\prime)
^ 2 dr
$$
These equations may be combined into a single 3rd order differential equation
as shown by Ray (1978) who may be said to have anticipated some
aspects of the results of Bartnik and Mckinnon. In any event Bartnik and
Mckinnon presented numerical evidence that there exist solutions with
$$
m (r) = O (r ^ 3); \ \ \ w (r) = 1 + O (r^ 2) \ \ \rm {at} \ \ r = 0
$$
and
$$
m(r) \sim m (\infty) - c ^ 2 /r ^ 3 ; \ \ \ w (r) \sim  \pm(1 - c/r) \ \ \rm
{at} \ \ r = \infty
$$
for some constant $c$.

The solutions are indexed by the number $k$ of zeros of $w(r); k = 1, \ 1, \ 2
\dots$.  The solutions have 3 regions.  Region I is an inner core.  Region III
is the asymptotic region where $F = 0 ({1 \over r ^ 3})$ and the metric
becomes Schwarzschildean.  The middle region has $w \approx 0$ and the
solution behaves rather like an abelian $U(1)$ Dirac monopole and the geometry
resembles the throat region of an extreme Reissner-Nordstrom solution.

The results of Bartnik and McKinnon have been confirmed by Kunzle and
Masood-ul-Alam (1990) and by Maison (private communication).

The question immediately arises:  are these solutions stable?  Bartnik and
McKinnon themselves felt that the cases $k \geq 3$ were unstable.  A stability
analysis was carried out by Straumann and Zhou (1990)and also (private
communications) by Maison.  These analyses show that these solutions are
unstable for {\bf {all}} values of $k$.  Presumably a small perturbation would
cause them either to collapse to form a black hole or (rather less likely) to
explode and dissipate.  Since they have no magnetic moment the expected hole
will be Schwarzschild like.  A noteworthy feature of the analysis of Straumann
and Zhou was that the configurations they considered were spherically
symmetric and yet time-dependent.  In other words Birkhoff's theorem is not
valid for Einstein-Yang-Mills unlike Einstein-Maxwell.

Thus Einstein-Yang-Mills admits unstable finite energy static non-singular
solutions.  We have seen above that the Einstein-Higgs equations do not.  What
about Einstein-Yang-Mills-Higgs?  It is physically clear that for small values
of $Gm/R$ where $R$ is a typical radius and $m$ a typical total energy the
effects of gravity on a 't Hooft-Polyakov monopole will be negligible.  Since
typically
$$
m \sim {4 \pi \eta \over e}
$$
$$
R \sim {1 \over e \eta}
$$
gravity will be negligible so long as:
$$
4 \pi G \eta ^ 2 \ll 1.
$$

If, on the other hand we consider a one parameter family of static solutions
labelled by the dimensionless number $4 \pi G \eta ^ 2$ (keeping $\lambda /e ^
2$ fixed) we might expect to encounter a critical value beyond which no static
solutions are possible because they will undergo gravitational collapse.  It
is also likely that the static family will already have become unstable at
some smaller value of $4 \pi G \eta ^ 2$.  The intuition one is drawing on
here is of course an analogy with the theory of white dwarf stars, the
critical value of $4 \pi G \eta ^ 2$ corresponding to the famous Chandrasehkar
limit.

As far as I know a detailed analysis of this situation has not been carried
out until recently. Miguel Ortiz in his Ph.D thesis has begun a numerical study and his results
confirm that for fixed $\lambda /{e^2}$ there is  a maximum value of $4 \pi G
\eta ^2 $ beyond which regular solutions without event horizons
cease to exist. This maximum value is about 2.5 in the Prasad-Sommerfeld
limit  and decreases as the quartic coupling constant $\lambda$ increases.
The metric at large distances appears to approach the Reissner-Nordstrom form
with an approximately minimal mass for a magnetic charge $g$, that is the
monopole appears to collapse as soon as the Cosmic Censorship allows. The
exterior gauge field and Higgs field  appear to approach a Wu-Yang like
configuration with the Higgs field being covariantly constant. This will be
described in sections 4 and five in more detail.

The basic equations for self-gravitating 't Hooft
Polyakov monopoles in
the spherically symmetric case were in fact  written down some time ago
by Perry, Van Nieuwenhuizen
and Wilkinson (1976). A variational principle was established but the
equations were not analysed in detail.  A qualitative physical discussion
along the lines indicated in the previous paragraph has been given by Frieman
and Hill (1987).   Some more exact information can possibly be obtained by
considering the generalizations of the Bogomolnyi bound in the gravitational
setting so we now turn to that topic.
\vfill
\eject
{\bf {3.  Bogomolnyi Bounds for Einstein-Yang-Mills-Higgs Initial Data}}

We shall consider time-symmetries initial data for simplicity, that is the
second fundamental form $K _ {ij}$ of the initial surface is assumed to
vanish.  In addition we assume that the non abelian electric field vanishes,
as well as the time component of the Higgs field's covariant
derivative.  Thus the Ricci scalar $R$ of the 3-metric $g _ {ij}$ satisfies
$$
R = 16 \pi G T _ {\hat 0 \hat 0}
\eqno (3.1)
$$
where
$$
T _ {\hat 0 \hat 0} = {1 \over 2}   B ^ 2 + {1 \over 2} (
D   \Phi ) ^ 2 + W (\Phi).
\eqno (3.2)
$$

Let us define the ``total amount of matter'' on the initial surface $\Sigma$
(assumed to be complete) by:
$$
M = \int _ \Sigma \ T _ {\hat 0 \hat 0} \sqrt( \ \  ^3 g) \ d^ 3 x
\eqno (3.3)
$$

Note that $M$ does not, in general, equal the ADM mass $m$ of the 3-metric $g _
{ij}$.  Even if it were the case that the data were such as to evolve to a
static solution a comparison of (3.3) and (2.5) shows that $M$ and $m$ cannot
be expected to coincide.  Another measure of the total energy of the matter in
a static spacetime would be the ``Killing Energy'' $E$ defined by
$$
E = \int _ \Sigma \sqrt( -g _ {00}) \ \ T _ {\hat 0 \hat 0} \sqrt {g} d ^ 3 x
\eqno (3.4)
$$
In general we have (when they are defined)
$$
M \not = E \not = M \not = m
$$

Now Bogomolnyi's original argument  may trivially be ``covariantised'' with
respect to spatial diffeomorphisms using the covariantly constant alternating
tensor $\epsilon_{ ijk}$ of the 3-metric $g _ {ij}$ (I prefer {\bf {not}} to
use tensor
densities and I am of course assuming that the initial surface is oriented).
Thus we have

{\bf {Theorem 2.  The total amount of matter $M$ of a time-symmetric initial
dates set for the $SU(2)$ Einstein-Yang-Mills-Higgs equations with non negative
potential $W (\Phi)$ is bounded below by}}
$$
M \geq g \eta
\eqno (3.5)
$$
{\bf {where $g$ is the asymptotic magnetic monopole moment}} .For a single
monopole $g={4 \pi} /e$.
Moreover equality in (3.5) implies that the covariant
Bogomolnyi equations hold.
$$
D _ i \Phi = \pm {1 \over 2} \epsilon _i \ ^ {jk}  \ F _ {jk}
\eqno (3.6)
$$
hold.

The existence of solution of (3.6) on a curved metric has been studied by
Floer (1987)).  Although of some mathematical interest the following result
shows that these solutions are never of relevance if the monopole
self-gravitates.

{\bf {Theorem 3.  Static solutions of the Einstein-Yang-Mills-Higgs equations
satisfying the Bogomolnyi equations (3.6) or equivalently saturating the
Bogomolnyi bound (3.5) do not exist.}}

{\bf {Proof}}  The Bogomolnyi equation (3.6) imply that the spatial components of
the stress tensor $T _ {ij} = 0$.  We can thus invoke our previous theorem 1.

It is known that to form an abelian black hole of ADM mass $m$ and
magnetic charge $g$ we must have
$$
m \geq {g \over \sqrt {4 \pi G}}
\eqno (3.7)
$$
(recall that we are using rationalized units for electromagnetic or Yang-Mills
fields).  Equality in corresponds to the Papapetrou-Majumdar metrics describing
the equipoise of an arbitrary number of extreme Reissner-Nordstrom magnetic
black holes. As mentioned in section 2  equation (3.7) shows that for fixed magnetic charge $g$ a 't
Hooft Polyakov monopole cannot collapse until its ADM mass $m$ satisfies $\sqrt
{4 \pi G} m \geq g$.  Since $m \sim M \sim \eta g$ we need $4 \pi G \eta ^ 2
\geq 1$ which agrees approximately with what has been found by Ortiz. It thus
seems very reasonable to expect that the configuration to which it
gives rise is similar, if not identical to, an extreme Reissner-Nordstrom
solution. We shall consider static solutions with horizons with horizons.  That we
in the next section.  Before doing so we remark that some information
about initial data for Einstein-Yang-Mills-Higgs has been obtained by Malec and
and Koc (1990) and Chmaj and Malec (1989).

If one is merely interested in the Yang-Mills equations in a gravitational
background it is possible to find a modified set of Bogomolnyi equations:
$$
D _ i   {\Phi} =   {B _ i} -   {\Phi} \nabla _ i U
\eqno (3.8)
$$
where $U$ is the Newtonian potential as defined by (2.6).  If the background
metric satisfies
$$
{\nabla  ^2} _ g  U = 0,
\eqno (3.9)
$$
then (3.8) implies the second order Yang-Mills equations in the background (see
Comtet (1980) and Comtet Forgacs and Horvathy (1984).
In general (3.9) will be incompatible with the Einstein-Yang-Mills
equations.  An interesting case for which (3.9) {\bf {is}} compatible with the
Einstein-Yang-Mills equations is when $\gamma _ {ij}$ is flat.  This gives the
Papapetrou-Majumdar metrics for which the left and right hand sides of (3.8)
are separately zero and $  B$ and $\Phi$ point in a constant direction
in internal space.  For more detail about these equations see Horvathy ((1987)
\vfill
\eject
{\bf {4.  Static solutions of the Einstein-Yang-Mills equations with Horizons}}

It has been known for many years that one has the abelian black hole solutions
with
$$ A = \tau _ 3 ({q \over r} dt + g \cos \theta d \phi)
\eqno (4.1)
$$
$$
ds ^ 2 = - \Delta dt ^ 2 + {dr ^ 2 \over \Delta} + r ^ 2 ( d \theta ^ 2 + \sin ^
2 \theta d \phi ^ 2)
\eqno (4.2)
$$
$$
\Delta = 1 - {2 Gm \over r} + G {(q ^ 2 + g ^ 2) \over 4 \pi r ^ 2}
\eqno (4.3)
$$
where $m$ is an arbitrary constant satisfying
$$
m \geq {1 \over \sqrt {4 \pi G}} (q ^ 2 + g ^ 2 ) ^ { 1 \over 2}
\eqno (4.4)
$$

If $\tau _ 3$ has a normalization such that
$$
\exp (4 \pi i \tau _ 3) = 1
 \eqno (4.5)
$$
we must demand that
$$
{eg \over 2 \pi} = 0, \ 1, \ 3, \ 5, \dots
\eqno (4.6)
$$
if $A$ is an $SO(3)$ connection and
$$
{eg \over 4 \pi} = 0, \  1, \  2, \ 3
\eqno (4.7)
$$
if $A$ is an $SU(2)$ connection.  Note that the $SO(3)$ case is only possible
because the presence of the horizon means that the singularity which would
otherwise result at $r = 0$ is hidden inside the horizon.  It cannot occur if
there are no horizons.

Note that (4.1) will always yield a spherically symmetric energy-momentum
tensor although it is not spherically symmetric as an $SU(2)$ connection unless
$$
{eg \over 4 \pi} = 1
\eqno (4.8)
$$

This latter case corresponds to $w = 1$ in (2.11).  Some authors prefer to use
a different gauge.  Let
$$
x = r (\sin \theta \cos \phi , \sin \theta \sin \phi, \cos \theta)
$$
then the connection
$$
e \tilde {A} ^ 1 = {dx ^ 2  \ \wedge \ dx ^ 3 \over r ^ 2} (w - 1) \ \ \rm {etc}
\eqno (4.9)
$$
is gauge equivalent to (2.11) with $a(r) = 0$.  Thus the Reissner-Nordstrom
metric with $q = 0$ and $\tilde A$ given by (4.9) with $w = 0$, whence (4.8)
represents the simplest purely magnetic $SU(2)$ solution.  This solution will
extend trivially to a solution of the Einstein-Yang-Mills-Higgs equations if
one appends the covariantly constant Higgs field
$$
\Phi ^ i =\eta { {x ^ i} \over r }
\eqno (4.10)
$$

The resulting solution is said to satisfy the Wu-Yang ansatz.  Of course in the
Abelian gauge it reduces to
$$
eA = 4 \pi \tau _ 3 \ \cos \theta d \phi
\eqno (4.11)
$$
$$
\Phi = \eta \tau _ 3
$$
$$
ds^2 = - (1 - {2Gm \over r} + {4 \pi G \over e ^ 2r^2} ) dt ^ 2 + (1 - {2Gm
\over r} + {4 \pi G \over e ^ 2 r^2} ) ^ {-1} dr ^ 2
\eqno (4.12)
$$
$$
+ r ^ 2 (d \theta ^ 2 + \sin ^ 2 \theta d \phi ^ 2 )
\eqno (4.13)
$$

Until the work of Bartnik and McKinnon it had long been felt that these abelian
solutions would be the only solutions - rather by analogy with the No-Hair
Theorems for the Einstein-Maxwell equations.  However recent work has clearly
indicated this conjecture to be false.  Kunzle and Masood-ul-Alam (1990),
Bizon (1990) , and
Volkov and Gal'tsov (1989) have shown that there exist analogues of the Bartnik and
McKinnon solutions with horizons.  As one might expect these are unstable, as
shown by Straumann and Zhou (1990). The no hair conjecture in its naive form thus
fails.  In the spherically symmetric case however Gal'tsov and Ershov (1990)
have argued that as long as there is a net Yang-Mills charge measurable at
infinity, i.e. that $w ^ 2 \not = 1$ at infinity then the abelian solutions are
unique.

What about the stability of the abelian solutions?  This was studied some time
ago by Lohiya (1982).  Following the analogous analysis of singular
monopoles in flat spacetime.  The stability is determined by the large distance
behaviour of the fields in a manner described by Coleman (19839 and Brandt and
Neri (1979).
The analysis shows that the purely magnetic solutions are unstable stable if
the connection is topologically trivial restricted to a 2-sphere at infinity.
This means that all $SU(2)$ connections i.e.
$$
{eg \over 4 \pi} \in Z
$$
are unstable.  Of the remaining $SO(3)$ connections only the lowest one with
$$
{eg \over 2 \pi} = 1
$$
is stable.

A sufficient condition for instability of the electric solutions is that the
electric charge $q$ exceeds $3e/2$.  Again this is consistent with the flat
space results.

The conclusion would seem to be that in the absence of Higgs fields regardless
of uniqueness the non abelian-Einstein-Yang-Mills solutions are not of very
much physical interest.  It is therefore appropriate to turn to the case when
Higgs fields are included.
\vfill
\eject
{\bf {5.  Solutions of the Einstein-Yang-Mills Higgs Equations with Horizons}}

The obvious first remark is to recall that solutions always exist if the Higgs
field is
covariantly constant.  Choosing a gauge in which the direction of $\Phi$ is
everywhere the same in internal space we see that Yang-Mills potentials
associated with rotations about that direction satisfy the abelian equations
and thus must belong to the Reissner-Nordstrom family described earlier.  Of
course the stability results of Lohiya do not necessarily apply now because the
Higgs mechanism might well stabilize these Dirac-type monopoles (for
sufficiently large Higgs mass) against instabilities in the non-abelian
directions. To my knowledge this has not been looked at by anybody in detail.

In the case that the electric charge vanishes
the results
of Ortiz suggest that gravitational collapse of a 't Hooft-Polyakov monopole
will result in an exterior field in which the Higgs field is covariantly
constant and given by (4.10), and the gauge field by (4.9) and $w=0$. This
solution was originally written down by Cho and Freund (1975) and Bais and Russell
(1975). As mentioned earlier these correspond to the Reissner Nordstrom solutions
with $q=0$ and $g= {{4 \pi} \over e}$. My conjecture is that these are indeed
classically stable. Moreover quantum mechanically they should evolve by Hawking
evaporaation to the extreme, zero temperature state. Such objects should behave
like stable solitons and have been studied extensively by Hajicek and his
collaborators from that point of view. Thus if $4 \pi G \eta ^2$ is large enough
the monopole problem of cosmology is in fact a primordial black-hole monopole
problem. In fact it it seems rather likely that 'tHooft-Polyakov monopoles will
be unstable for values of $4 \pi G \eta^2 $ which are somewhat smaller than the
maximum allowed value.

An important  question now arises:  are there any other solutions? For eaxmple
are there any electrically charged solutions in which the electric charges are
associated with the broken $SU(2)$ generators for example ?
Experience and intuition
based on both the physical ideas behind the Higgs mechanism (charge should be
screened) and the non-hair properties of black holes would have suggested until
very recently that the answer is no.  At present however the answer is less
clear because of two developments.  The first is the
Bartnik-McKinnon-Bizon-Kunzle-Masood-ul-Alam - Vokkov-Gal'tsov solutions.  The
second is the issue of fractional charges raised by  Krauss and
Wilczek (1990), see Preskill and Krauss (1990) and Preskill (1990).
Even in the simpler Abelian-Higgs model the situation is not entirely
clear.  For that reason I will discuss what is known in that case.

The
simplest question to ask is are there static solutions of the Einstein-Higgs
equations with horizons?  By static I mean that not only is the metric static
but that the complex Higgs field which I shall now call $\phi$ is
strictly independent of time.  If
one doesn't make that assumption one might expect to find shells of
``Q-matter'' surrounding a black hole.  It is generally expected that as long
as $W(\phi)$ is positive with an absolute minimum at $|\phi | = \eta$ then the
only solution must have $\phi =$ constant, with the constant real with no loss
of generality.  If $W (\phi)$ vanishes it is easy to establish this result.  If
$W(\phi)$ is{ \bf {convex}} it is also possible to establish this result using a
``Bochner Identity''. Suppose, more generally, that a field $\phi ^ A
(x)$ takes its values in some riemannian  target manifold $N$ with metric $G_ {AB} ((\phi)$ and
potential function $W (\phi)$.  The Bochner identity tells us that:
$$
({1 \over 2} G _ {AB} {\partial \phi ^ A \over \partial x ^ \alpha} {\partial
\phi ^ B \over \partial x ^ \beta} g ^ {\alpha \beta} ) ^ {; \mu} _ {; \mu} =
\phi ^ {A ; \alpha ; \beta} \phi ^ {B} \ \ _{ ;  \alpha ; \beta} G _ {AB}
$$
$$
+ \phi ^ {A;\alpha} ( G _ {AB} R _ {\alpha \beta} - g _ {\alpha \beta} R _
{ACBD} \phi ^ C _ {; \mu} \ \phi ^ D _ {; \nu} \ g ^ {\mu \nu} ) \phi ^ {B ;
\beta}
$$
$$
+ \phi ^ {A ; \alpha} ( \phi ^ {B ;\beta}\ _ {;\beta} ) _ {; \alpha} G_{AB}
\eqno (5.1)
$$
where all covariant derivatives are covariant with respect to the spacetime
metric $g _ {\alpha \beta}$ and the target-space metric $G _ {AB}$ in the manner
described by Misner (1978).  The field equations are:
$$
\phi ^ {A ; \beta} \ \ _ {; \beta} = G ^ {AB} \nabla _ B W
\eqno (5.2)
$$
and
$$
R _ {\alpha \beta} = 8 \pi G [ G _ {AB} \phi ^ A _ {; \alpha} \phi ^ B _
{; \beta} + g _ {\alpha \beta} W (\phi)]
\eqno (5.2)
$$

If one integrates (5.1) over the region exterior to the black hole and confined
within 2 spacelike surfaces $\sum _ 1 $ and $\sum _ 2$ such that $\sum _ 1$ is
the time translation of $\sum _ 2$ and uses the field equations and the boundary
conditions:
$$
\phi ^ A _ {; \alpha} \ \  l ^ \alpha = 0
$$
on the horizon and $\phi ^ A \rightarrow$ constant at $\infty$ and  where $l ^
\alpha$ is the null generator of the horizon one obtains the following

{\bf {Theorem 4  (Scalar No Hair theorem)  There are no non-trivial static
scalar fields on a static black hole solution of the Einstein-Higgs equations
for which $W (\phi) \geq 0$, and the sectional curvature of the target manifold
is non positive and $W (\phi) _ {;A;B}$ is non negative}}

Note that unlike Cor. 1 of Theorem 1 we need a stronger assumption on $W (\phi)$
and $G _ {AB}$.  If $G _ {AB}$ is flat and $\phi ^ A$ takes its value in a
vector space we could have used the simpler identity:
$$
(\phi ^ A G _ {AB} \phi ^ B _ {; \alpha} ) ^ {; \alpha} = \phi ^ A _ {; \alpha}
G _ {AB} \phi ^ B _ {; \beta} \ \  g ^ {\alpha \beta}
\eqno (5.4)
$$
and the field equation (5.2) one obtains

{\bf {Theorem 5:  There are no non-trivial static scalar fields on a static
black hole solution of the Einstein-Higgs equations for which $\phi ^ A W
(\phi) _ {;B} \geq 0$}}

Remark:  The proof of theorems 4 and 5 also applies to the case where infinity
is replaced by a cosmological event horizon.

Neither theorem 4 nor theorem 5 applies to even the simplest case of a single
real scalar field $\phi$ with potential
$$
W (\phi) = \lambda (\phi ^ 2 - \eta ^ 2) ^ 2
\eqno (5.5)
$$
$\lambda > 0$.  Thus for non-linear field equations of this type the no-hair
conjecture remains - to use a standard term in Scots law -\lq \lq Not Proven
\rq \rq although
there is some suggestive work by Sawyer (1977)  and Brumbaugh (1978) .

Let us turn to the work of Adler and Pearson (1978).  They assume spherical symmetry
and the Einstein-Maxwell-Higgs equations, with a complex scalar $\phi$.  They
assume that  that  there is only an electric field
present and a gauge exists in which:

(1) $\phi$ is independent of time and real

(2) $A = A _ 0 dt$ with $A _ 0$ everywhere bounded

Actually their assumptions are unnecessarily restrictive and their arguments
both incomplete and in part wrong.  We shall assume to begin with that
$$
A _ \mu = A _ 0 dt
\eqno (5.6)
$$
with

(1)  $\phi, \ A _ 0$  independent of time

(2) $A _ 0 \ \rightarrow 0$ at $\infty$

(3)  the one form $A_0 dt= A_{\mu} dx^{\mu}$ has
bounded ``length'.'
That is $g ^ {\mu \nu} A _ \mu  A  _ \nu <
\infty$ on the horizon.

The equation for $A _ 0$ is
$$
- \nabla _ j (V ^ {-1} \nabla ^ j A _ 0) + e ^ 2 | \phi| ^ 2 A _ 0 = 0
\eqno (5.7)
$$
where $\nabla$ is taken with respect to the 3-metric $g _ {ij}$. We
have dropped the assumption that the metric is
spherically symmetric and that $\phi$
is real.

If one multiplies (5.7) by $A _ 0$ and integrates over a surface of constant
time one obtains:
$$
\int _ \Sigma \sqrt {g} (V ^ {-1} e ^ 2 | \phi|^2 (A _ 0) ^ 2 + V ^ {-1} (
\nabla ^ i A _ 0 \nabla ^ j A _ 0) g _ {ij})
$$
$$
= \int _ {\partial \Sigma} ( V ^ {-1} A _ 0 \nabla ^ j A _ 0 ) d \sigma _ j
\eqno (5.8)
$$

Now if $|\phi| \rightarrow \eta$ and $V \rightarrow 1$ at infinity solutions
of (5.7) at infinity to like ${1 \over r} exp \pm e \eta r $.  Thus if $A _
0$ is to be bounded it must fall to zero exponentially and the boundary term
at infinity (5.9) will vanish.  On the other hand if the field strength $F _
{i0} = \partial _ i A _ 0$ is to have bounded scalar invariant on the horizon
we require that $V ^ {-2} (\nabla _ j A _ 0) ( \nabla ^ j A _ 0)$ should be
bounded near the horizon.  Now if in addition $A _ \mu A _ \nu g ^ {\mu \nu}$
is to be bounded we require that $A _ 0$ vanishes at least as fast as $V$ at
the horizon and so the boundary term at the horizon in (5.9) must vanish.

We have thus established the following:

{\bf {Lemma 1}}  There are no regular time independent electrostatic fields
with time independent vector potentials and Higgs field which are bounded with
bounded length $A _ \mu
g ^ {\mu \nu} A _ \nu$ around a static black hole.

Unfortunately  lemma 1 is not sufficient to establish that there can be
no time independent electrostatic fields around a black hole because it is
not clear that there should exist a global gauge in which  the vector
potentials and Higgs fields are both  time independent and bounded. In the
usual electromagetic case without symmetry breaking the potential $A_\mu$ cannot
in fact be cast in a such a gauge. Thus it is necessary to investigate the case
when either the gauge variant fields $A_\mu$ and $\phi$ vary with time or do
not fall off at infinity. To my knowledge this has not been done.

Even if one assumes that the electromagnetic field vanishes and that the
Higgs field is time independent and bounded and
even if one assumes further that it is real I know of no rigorous proof that
it must be
constant in the case that the potential $W$  has the (non-convex) form (5.5)..
The argument given
be Adler and Pearson for example appears to be incorrect. Although the no-hair
property seems very plausible physically it is clear that much remains to be
done to establish it rigorously even in the abelian case with symmetry breaking
let alone in the non-abelian case.
\vfill
\eject
{\bf {6.  Global Monopoles and Black Holes}}

Barriola and Vilenkin(1989) have pointed out that the gravitational
field
of a global monopole has some interesting properties.  Far from the core one
has
$$
\Phi ^ i \simeq \eta X ^ i / r
\eqno (6.1)
$$
$$
T ^ o _ o \simeq \eta ^ 2 1 / r ^ 2
\eqno (6.2)
$$
$$
T ^ r _ r \simeq \eta ^ 2 1 /r ^ 2
\eqno (6.3)
$$
$$
T ^ \theta _ \theta = T ^ \phi _ \phi \simeq - \eta ^ 2 / r ^ 2
\eqno (6.4)
$$
with asymptotic metric
$$
ds ^ 2 \simeq - dt ^ 2 + {dr ^ 2 \over 1 - 8 \pi G \eta ^ 2} + r ^ 2 (d \theta
^ 2 + \sin ^ 2 \theta d \phi ^ 2)
\eqno (6.5)
$$
This metric is not asymptotically flat but rather asymptotically the product
of a flat time direction (i.e. the Newtonian potential tends to zero) with a
3-dimensional cone over $S ^ 2$ with solid angular deficit $32 \pi ^ 2 G
\eta ^ 2$.  For a non singular (but infinity total energy) $\Phi ^ i$ must
vanish at $r = 0$, and the metric acquires some corrections.  nevertheless in
the ``$\sigma$-model approximation'' in which $\Phi ^ i$ always remains in the
global minimum of $W (\Phi)$ one may replace the $\simeq$ in (6.1) - (6.5) by
= signs.

They are exact solutions of the Einstein equations with $\sigma$-model
source.  Moreover one may consider in addition a black hole.  Then (6.1) -
(6.4) continue to hold as equalities and (6.5) is replaced by
$$
ds ^ 2 = - (1 - {2Gm \over r} ) dt ^ 2 + {dr ^ 2 \over (1 - 8 \pi G \eta ^
2) (1 - {2Gm \over r})} + r ^ 2 (d \theta ^ 2 + \sin ^ 2 \theta d \phi ^ 2)
\eqno (6.6)
$$
The metric (6.6) (which was worked out by myself and Fernando Ruiz-Ruiz)
thus represents a global monopole inside a black hole.  Unfortunately
however, as pointed out by Goldhaber (1989), the global monopole is
likely to be unstable against a sort of angular collapse in which all the
$\Phi$ field energy becomes concentrated along a line defect or string
leaving the point defect at $r = 0$.  This string has an energy per unit
length of $4 \pi \eta ^ 2$.  The analysis of Goldhaber is consistent with
the work of a number of people on defects in liquid crystals which are
modelled using the a free-energy functional [the ``Frank Oseen free energy
in the one constant approximation] which particle physicists would refer to
as a $\sigma$-model action and mathematicians as an harmonic map energy
functional.  Point defects have strings emerging from them which tend to the
zero thickness limit of the $\sigma$-model cosmic strings introduced by
Comtet and myself (1989). Despite this instability there continue to appear
preprints analyzing there properties and those of similar objects. An
interesting feature is that under some circumstances there can be repulsive
gravitational effects. In particular  Harari and Lousto (1990) have drawn attention to
a repulsive region near the core. A similar feature was found by Ortiz near the
core of a 't Hooft-Polyakov monopole. An interesting question to ask is whether
for large enough $4 \pi \eta ^2$ gravitational collapse is inevitable and what
is the critical value ? In effect this is a limiting case of the Ortiz problem
when $\lambda/{e^2} $ is large.

The gravitational field of an infinite straight
$\sigma$-model strings was given by myself and Comtet (1989).  What about that due to a
string emerging from a black hole?  Such a string would cause the black hole
to accelerate and so the appropriate solution (in the thin string
approximation) is the C-metric:
$$
ds ^ 2 = {1 \over A ^ 2 (x +y ) ^ 2} [ {d y ^ 2 \over F (y)}
 + {dx ^ 2 \over G (x)} + G (x) d \alpha ^ 2    -F(y)dt^2 ]
$$
where
$$
G (x) = - F (-x)
$$
$$
= 1 - x ^ 2 - 2GmAx^3; \ \ \ \ \ 0 < GmA < 1/\sqrt {27}
$$
$$
= 2GmA (x - x _ 2) (x - x _ 3) (x - x _ 4)
$$

I have labelled the 3 real roots of $G (x) \ x _ 2, \ x _ 3,  \  x _ 4$
in ascending magnitude ($x _ 2$ and $x _ 3$ are both negative and $x _ 4$ is
positive).

The range of the ``radial'' variable $y$ is
$$
-x _ 3 \leq y \leq - x _ 2
$$
with $y = |x _ 3|$ being an acceleration horizon and $y = |x _ 2|$ a black
hole horizon.  The range of the ``angular'' variable $x$ is $x _ 3 \leq
\alpha \leq x _ 4$.  The 2-surfaces $x = x _ 3$ and $x = x _ 4$ are axes of
symmetry for the angular Killing vector ${\partial \over \partial \alpha}$ .

In order to understand what the coordinates used it is helpful to consider the
case when the the mass parameter $m$ vanishes. Then the metric is flat and one
may transform to flat
inertial coordinates using the formulae:
$$
X^1  \pm iX^2 = {{(1-x^2)^{1 \over 2}} \over {A(x+y)}} \exp (\pm i \alpha)
$$

$$
X^3 \pm X^0 = {{(y^2-1)^{1 \over 2}} \over {A(x+y)}} \exp (\pm t )
$$

Evidently the coordinate singularity at $x=\pm 1$ is a rotation axis while
the coordinate singularity at $y= \pm 1$ corresponds to a pair of intersecting
null hyperplanes forming the past and future event horizons for a family of
uniformly accelerating worldlines. A similar interpretation may be given in the
case  that $m \neq 0$ but there is in addition a Black Hole horizon. A detailed
description was given by Kinnersley and Walker (1970)

If $0 \leq \alpha \leq \Delta \alpha$ there will be angular deficits:
$$
{\delta _4 \over 2 \pi} = {\Delta \alpha - \Delta \alpha _4 \over \Delta
\alpha _4} \ \ \ ; \ \ \ \ {\delta _ 3 \over 2 \pi} = {\Delta \alpha -
\Delta \alpha _ 3 \over \Delta \alpha _ 3}
$$
where
$$
\Delta \alpha _ 4 = {4 \pi \over |G ^ {\prime} (x_4 )|} \ \ \ \ \ ; \ \ \ \ \ \Delta
\alpha _ 3 = {4 \pi \over |G ^ {\prime} (x_3)|}
$$
Since (unless $mA = 0$) $\Delta \alpha _ 4 \not = \Delta \alpha _ 3$ it is not
possible to eliminate both of these by choosing $\Delta \alpha$.  One can
eliminate $\delta _ 3$ in which case the black hole is pulled along by a
string, or $\delta _ 4$ in which case it is pushed along by a rod.  In general
the net ``force'' on the hole is
$$
F = { \delta _ 4 \over 8 \pi G} - {\delta _ 3 \over 8 \pi G} =
{\Delta \alpha \over 4G} ({1 \over \Delta \alpha _ 4} - {1 \over \Delta \alpha _
3}) = {\Delta \alpha mA \over 8 \pi} (x _ 4 - x _ 3) ^ 2
$$

The black hole event horizon area ${\cal  A}$ is given by
$$
{\cal A} = {\Delta \alpha \over A ^ 2} {x _ 4 - x _ 3 \over (x _ 4 - x _ 2) (x _ 3 - x
_ 2)}
$$

The black hole horizon surface gravity $\kappa _ {BH}$ and the
acceleration horizon
surface gravity $\kappa_ R$ are given by:
$$
\kappa _ {BH} = GmA ^ 2 (x _ 3 - x _ 2) (x _ 4 - x _ 2)
$$
$$
\kappa _ R =GmA^2 (x _ 3 - x _ 2) (x _ 4 - x _ 3)
$$
If $GmA << 1$ one obtains:
$$
\kappa _ {BH} \sim {1 \over 4Gm}
$$
$$\kappa _ R \sim A
$$
$$
{\cal A} \sim 8 \Delta \alpha G^2m^2
$$
$$
F \sim \Delta \alpha {mA \over 2 \pi}
$$
whence
$$
F \simeq {{\cal A} \kappa_{BH} \over 8 \pi G} \ \ . \ \ \kappa _ R
$$
which is equivalent to Newton's second law of motion.  However for finite $mA$
one does not obtain such a simple expression.  This is perhaps not surprising
since if $mA$ is not small the Schwarzschild radius of the black hole is
comparable with the radius of curvature of its world line.  Nevertheless it
would be nice to understand the relation between mass, acceleration and force
in this non-linear situation.  Some attempts in this direction, which also
relate to black hole thermodynamics were made by Aryal, Ford and Vilenkin
(1986),( see also Martinez and York (1990).
Note that Aryal te al. use the  representation of
accelerating black hole metrics in terms of Weyl-metrics using the ``rod
representation'' of Schwarzschild (Israel and Khan (1964).  The relationship between
this picture and the C-metric including the co-ordinate transformation between
the finite rod plus semi-infinite rod (each of mass per unit length ${1 \over
2}$) and the C-metric form quoted above may be found in (Godfrey (1972) see
also (Bonnor (1983,1990).More about strings and black holes may be found in
Chandraskhar and Xanthopouls (1989).
\vfill
\eject
{\bf {7.  Black Hole Monopole Pair-Production}}

In the quantum theory we know that charged particle anti-particle pairs may be
created by a sufficiently strong electric field - a process sometimes called
the Schwinger Process.  It is plausible that magnetic monopoles should
similarly be created by strong magnetic fields.  This process was investigated
in Yang-Mills-Higgs theory by Affleck and Manton (1982) using instanton
methods.  The use of instanton methods to calculate the rate of production by
the Schwinger process is discussed in (Affleck, Alvarez and Manton).

Some time ago I suggested that the same process should occur in quantized
Einstein-Maxwell theory (Gibbons 1986). The idea has recently been taken up
again by
Strominger and Garfinkle (1990).  Since {\bf {pure}} Einstein-Maxwell theory
has invariance under the duality transformation
$$
F _ {\mu \nu} \rightarrow (\exp \ \alpha \star) F _ {\mu \nu}
$$
where $\star$ is the Hodge star operation on   2-forms.
There is no invariant
distinction between electric and magnetic, so let us concentrate on the purely
magnetic case.  In any event it is this case which is physically most
interesting in  more realistic models.

To begin we need to model a strong magnetic field coupled to gravity.  The
natural choice is the Melvin solution which represents an infinitely long
straight self-gravitating Faraday flux tube in equilibrium, the gravitational
attraction being in equipoise with the transverse magnetic pressure (Melvin
(1964).
The metric is:
$$
ds^2 = (1+ \pi G  B^2 \rho^2 )^2 (-dt^2 +dz^2 +d \rho ^2 ) + {\rho ^2 d
\phi ^2 } (1 +  \pi G B^2 \rho ^2 )^{-2}
$$
The magnetic field is given by:
$$
F={{B \rho d \rho \wedge d \phi } \over {(1 + \pi G B^2 \rho ^2 )^2}}
$$

The Melvin solution posesses a degress of uniqueness.  For example Hiscock
(1981) has
shown

{\bf
Theorem:  The only axisymmetric, static solution of the
Einstein-Maxwell field equations without an horizon which is
is asymptotically Melvin is in fact the Melvin Solution.
}

In fact Hiscock also allows for a neutral or electrically charged black hole as
well.

I myself can show:

{\bf
Theorem:  The only translationally invariant, static solution of the
Einstein-Maxwell field equations without horizon which is
asymptotically  Melvin is in fact the Melvin solution.
}

Proof: assume the metric is static and has reflection invariance with respect
to the $z-$direction. These two assumptions may easily be justified. The metric
takes the form
$$
ds^2=-V^2 dt^2 +Y^2 dz^2 +g_{AB}dx^A dx^B
$$
with $A=1,2$. The field equations are:
$$
\nabla ^A (VY \nabla _A  \ln (V/Y) )=VY8 \pi G( T_{ \hat z  \hat z } +
T_ { \hat 0 \hat 0 }   )
$$
$$
\nabla ^A (VY \nabla _A \ln (VY)) = VY 8 \pi G T ^A  _A
$$
$$
V ^{-1} \nabla _A \nabla _B V + Y^{-1} \nabla _A \nabla _B Y = K g_{AB} -8 \pi
(T_{AB} - {1 \over 2 } g_{AB}(T ^A _ A + T _{\hat z \hat z } + T _{ \hat 0
\hat 0  } ) )
$$
where $ K $ is the Gauss curvature of the 2-metric $g_{AB}$.
The electromagnetic field is assumed to be of the form:
$$
F= {1 \over 2 } F _{AB} dx ^A dx ^B .
$$
It follows that $T_ { \hat 0 \hat 0 } + T _{ \hat z \hat z } =0$ and hence:
$$
\nabla ^A ( VY \nabla _A (V/Y))=0.
$$
Now $V/Y$ tends to one at infinity (asymptotic boost invariance) and so we may
invoke tha Maximum Principle to show that $V=Y$ everywhere. Thus the metric
  must be boost invariant.It now follows that
$$
\nabla_A \nabla _B V = f g_{AB}
$$
for some scalar $f$. Thus
$$
K^A = \epsilon ^{AB} \nabla _B V
$$
is a Killing vector field of the 2-metric $g_{AB}$ and since
$K^A \partial _A V =0$
it is also a Killing vector field of the entire 4-metric.
It is not difficult to see
that this Killing vector field  corresponds to rotational symmetry of the
solution.

The argument just given will generalise in various ways to cover some
other stress  tensors and as mentioned above the staticity asummption and the
assumption that $g_{\alpha z }=g_{zz} \delta _{\alpha z }$ is not difficult to
justify using standard methods on the global theory of balck holes.
Interestingly however it does not seem to be possible to show using this method
that the metric of a local cosmic string must be axisymmetric. Even in flat
spacetime this seems to be a very difficult problem, i.e. to show that all
time independent
Nielsen-Olesen vortex solutions of the abelian Higgs model ( in the
non-supersymmetric case) must have axial symmetry.
Having established the credentials of the Melvin solution as uniquely suitable
model of a static magnetic field in general relativity we turn to looking for
instanton solutions representing the creation of a black hole monopole
anti-monopole pair. If there were no external magnetic field the obvious
candidate instantons would be the magnetically charged C-metric for which
$$
G(x) = 1-x^2 -2GmA x^3 - G(g^2/4 \pi ) A^2 x^4 .
$$
However this has nodal singularities. In fact since  the metric is boost
invariant it has zero ADM mass and thus it cannot be regular by the positive
mass theorem generalised to include apparent horizons. However it was pointed
out by Ernst ((1976) that the nodal singularity may be eliminated if one appends a
suitable magnetic field. The resulting metric is of the same form as (6.7) but
the first three terms are mutiplied by and the last term divided
by the
factor:
$$ (1+GBgx/2)^4 .
$$
If $m=0=g=A$ we get the Melvin solution but the limit must be taken carefully.
The nodal singularity may be eliminated if $B$ is chosen so that
$$
G^{\prime}(x_3) /(1+GBgx_3/2)^4 \ \ + G^{\prime} (x_4)/(1+Ggx_4/2)^4 =0.
$$
Where $x_3$ and $x_4$ are two larger roots of $G(x)$ and we assume now that
there are 4 roots. The smallest root $x_1$ is thus inside the acceleration
horizon.
This equation may be regarded as an equation for $B$ the magnetic field
necessary to provide the force to accelerate the magnetically charged black
hole. It is difficult to find an explicit solution in terms of $g$, $m$ and $A$
except when $GmA$ is small in which case one finds the physically sensible
result:
$$
gB \approx mA .
$$
In order to obtain an instanton which is regular on the Riemannian section
obtained by allowing the time coordinate $t$ to be pure imaginary it is
necessary that the $\tau = i t $ is perodic with period given by the surface
gravity. This leads to the condition that
$$
G^{\prime}(x_2) + G^{\prime}(x_3)=0.
$$
It appears that the the only way to satisfy this condition is to set:
$$
m=|g|/{\surd (4 \pi G})
$$
Note that this equation {\bf does not } mean that the horizons have vanishing
surface gravity as I mistakenly asserted in (1986).It is not difficult to see
that the topology of the Riemann section is $S^2 \times S^2 $ with a point
removed. In fact topologically one can obtain this manifold from $R^4$, which
is the topology of the Melvin solution, by surgery along an
$ S^1$. That is by cutting out a neighbourhood of a circle which has topology
$D^3 \times S^1$ with boundary $ S^2 \times S^1$ and replacing by $S^2 \times
D^2$ which has the same boundary. This surgery is also what is needed to
convert
$R^3 \times S^1$ to $R^2 \times S^2$ i.e. to convert a manifold with the
topology of "Hot Flat Space" to that with the topology of the Riemannian
section of the Schwarzschild solution.

The existence of this
instanton would seem to be rather important. It seems to imply for
example that it would be
{\bf
inconsistent} not to consider the effects of black hole monopoles since given
strong enough magnetic fields they will be spontaneously created. Once they are
created they should evolve by thermal evaporation to the extreme zero
temperature soliton state. Another reason why I believe that this process is so
important is that it seems to show that while one may have one's doubts about
the effects of wormholes because of the absence of suitable solutions of the
classical equations of motion with positive definite signature, the solutions
described here do indicate that some sort of topological fluctuations in the
structure of spacetime {\bf must} be taken into account in a satisfactory
theory of gravity coupled to Maxwell or Yang-Mills theory.
\magnification= \magstep 1
\openup 1 \jot
\vsize=23.5true cm
\hsize=16true cm
\nopagenumbers
\topskip=1truecm
\headline={\tenrm\hfill\folio\hfill}
\raggedbottom
\abovedisplayskip=3mm
\belowdisplayskip=3mm
\abovedisplayshortskip=0mm
\belowdisplayshortskip=2mm
\normalbaselineskip=12pt
\normalbaselines
\eject

\centerline {References}
\vskip 2.0cm
\item  S L Adler and R P Pearson, No Hair theorem for the Abelian Higgs
and Goldstone models.  Phys Rev{ \bf {D18}}, 2798-2803 (1978)
\item   I K Affleck, O Alvarez and N S Manton, Pair production at strong
coupling in weak external fields.  Nucl. Phys{ \bf {B197}} 509-519 (1982)
\item   I K Affleck and N S Manton  Monopole pair production in a magnetic
field.  Nucl Phys {\bf {B194}} 38-64 (1982)
\item   F J Almgren and E H Lieb, Counting Singularities in liquid
crystals.  Proc. IXth International Congress on Mathematical Physics, eds B
Simon, A Truman and I M Davies, Adam Hilger 1989
\item  F Almgren and E Lieb, Singularities of energy minimizing maps form
the ball to the sphere:  examples counter examples and bounds, Ann of Math.
{\bf {128}} 483-430 (1988)
\item   M Aryal, L H Ford and A Vilenkin, Cosmic Strings and Black Holes,
Phys Rev {\bf {D34}} 2263-2266 (1986)
\item   A Ashtekar and T Dray, On the Existence of solutions to Einstein's
Equation with Non-Zero Bondi News.  Commun. Math. Phys {\bf {79} }581-589 (1981)
\item  F A Bais and R J Russell, Magnetic-monopole solution of the
non-Abelian gauge theory in curved spacetime.  Phys Rev {\bf {D11}} 2692-2695
(1975)
\item   R Bartnik and J McKinnon, Particle like solutions of the
Einstein-Yang-Mills equations.  Phys Rev Lett {\bf {61}} 141-144 (1988)
\item   M Barriola and A Vilenkin, Gravitational field of a global
monopole. Phys. Rev. Lett {\bf {63}} 341-343 (1989)
\item  J Bicak, The motion of a charged black hole in an
electromagnetic field. Proc. Roy. Soc. {\bf {A371}} 429-438 (1980)
\item  W B Bonnor, The sources of the vacuum C-metric. Gen. Rel. Grav.
{\bf{15}} 535-551 (1983)
\item  W B Bonnor, The C-metric in Bondi's coordinates. Class. Quant. Grav.
{\bf {7}} L229-L230 (1990)
\item   P Bizon, Colored Black Holes.  Phys. Rev. Letts {\bf {64}}
2844-2847 (1990)
\item  P J Braam, A Kaluza-Klein approach to hyperbolic
three-manifolds.  Enseign. Math {\bf {34}} 275-311 (1985)
\item  R A Brandt and F Neri, Stability Analysis for Singular
Non-Abelian  Magnetic monopoles.
Nucl. Phys. {\bf {B161}} 253-282 (1979)
\item  P B Breitenlohner, G W
Gibbons and D Maison,
4-dimensional Black Holes from
Kaluza-Klein Theory. Commun. Math.
Phys. {\bf{120}} 295-334 (1988)
\item   H Brezis, J M Coron and E Lieb, Harmonic Maps with defects.
Commun. Math. Phys {\bf {107}} 649-705 (1986)
\item   B E Brumbaugh, Nonlinear scalar field dynamics in Schwarzschild
geometry.  Phys. Rev. {\bf {D18}} 1335-1338 (1978)
\item  S Chandrasehkar and B C Xanthopoulos two Black Holes attached
to strings.  Proc. Roy. Soc. {\bf {A423 }} 387-400 (1989)
\item   T Chmaj and E Malec, Magnetic monopoles and gravitational collapse.
Class and Quantum Grav. {\bf {6}} 1687-1696 (1989)
\item   Y M Cho and P G O Freund, Gravitating 't Hooft monopoles.
Phys. Rev. {\bf {D12}} 1588-1589, (1975)
\item  S Coleman in \lq\lq The Unity of Fundamental interactions\rq\rq
ed. A Zichichi (Plenum, New York ) (1983)
\item   A Comtet, Magnetic Monopoles in curved spacetimes.  Ann. Inst. H
Poincare {\bf {23}} 283-293 (1980)
\item   A Comtet, P Forgacs and P A Horvathy, Bogomolnyi-type equations
in curved spacetime.  Phys. Rev {\bf {D30}} 468-471 (1984)
\item   A Comtet and G W Gibbons, Bogomol'nyi Bounds for Cosmic
Strings.   Nucl. Phys. {\bf{B299}} 719-733
(1989)
\item A D Dolgov, Gravitational Dipole. JETP Lett. {\bf{51}} 393-396 (1990)
\item   T Dray, On the Asymptotic Flatness of the C Metrics at Spatial
Infinity.
Gen.Rel.Grav. {\bf {14}} 109-112 (1982)
\item   T Dray and M Walker, On the regularity of Ernst's generalized
C-metric.  L.I.M.P. {\bf {4}} 15-18 (1980)
\item  F J Ernst, Black holes in a magnetic universe. J.M.P. {\bf{17}}
54-56 (1976)
\item  F J Ernst, Removal of the nodal singularity of the C-metric.
J.M.P. {\bf {17}} 515-516 (1976)
\item   F J Ernst, Generalized C-metric.  J.M.P. {\bf {19}} 1986-1987
(1978)
\item  F J Ernst and W J Wild, Kerr black holes in a magnetic
universe.  J.M.P. {\bf {17}} 182-184 (1976)
\item  A A Ershov and D V Gal'tsov, Non Existence of regular monoples and
dyons in the SU(2) Einstein-Yang-Mills theory. Phys. Letts. {\bf 150A} 159-162
(1990)
\item   J A Frieman and C T Hill, Imploding Monopoles.  SLAC-PUB-4283
Oct. 1987 T/AS
\item   A Floer, Monopoles on asymptotically Euclidean manifolds.  Bull.
AMS {\bf {16}} 125-127 (1987)
\item   D V Gal'tsov and A A Ershov, Non-abelian baldness of coloured
black holes.  Phys. Letts {\bf {A138}} 160-164 (1989)
\item   D Garfinkle and A Strominger, Semi-classical Wheeler Wormhole
Production.  UCSB-TH-90-17
\item  G W Gibbons
Non-existence of Equilibrium
Configurations of Charged Black
holes. Proc. Roy. Soc. {\bf{A372}}
535-538 (1980)
\item   G W Gibbons, Quantised Flux-Tubes in Einstein-Maxwell theory and
non-compact internal spaces, in Fields and Geometry Proc. of XII Karpac Winter
School of Theoretical Physics 1986, ed A Jadczyk, World Scientific
\item   B B Godfrey, Horizons in Weyl metrics exhibiting extra
symmetries.
G.R.G. {\bf {3}} 3-15 (1972)
\item   A Goldhaber, Collapse of a Global Monopole.  Phys. Rev. Letts {\bf
{63}} 2158(c) (1989)
\item   {Gu} Chao-hao, On Classical Yang-Mills Fields.  Phys. Rep. {\bf
{80}} 251-337 (1981)
\item   P Hajicek, Wormhole solutions in the Einstein-Yang-Mills-Higgs
system.  I General theory of zero-order structure.  Proc. Roy. Soc A {\bf {386}}
223-240 (1983)
\item  P Hajicek, Wormhole solutions in Einstein-Yang-Mills-Higgs system
II Zeroth-order structure for $G = SU(2)$. J. Phys. {\bf {A16}} 1191-1205 (1983)
\item  P Hajicek, Classical Action Functional for the system of fields
and wormholes.  Phys. Rev. {\bf {D26}} 3384-2295 (1982)
\item   P Hajicek, Generating functional $Z _ 0$ for the one-wormhole
sector.  Phys. Rev. {\bf {D26}} 3396-3411 (1982)
\item  P Thomi, B Isaak and P Hajicek, Spherically Symmetric Systems of
Fields and Black Hole. I Definition and properties of Apparent Horison.  Phys.
Rev. {\bf {D30}} 1168-1171 (1984)
\item   P Hajicek, Spherically symmetric systems of fields and black
holes.  II Apparent horizon in canonical formalism.  Phys. Rev. {\bf {D30}}
1178-1184 (1984)
\item   P Hajicek, Spherically symmetric systems of fields and black
holes.  III Positively of enemy and a new type of Euclidean action.  Phys. Rev.
{\bf {D30}} 1185-1193 (1984)
\item   P Hajicek, Spherically symmetric systems of fields and black
holes.  IV No room for black hole evaporation in the reduced configuration
space?  Phys. Rev. {\bf {D31}} 785-795 (1985)
\item   P Hajicek, Spherically symmetric systems of fields and black
holes.  V  Predynamical properties of causal structure.  Phys. Rev. {\bf {D31}}
2452-2458 (1985)
\item   P Hajicek,  Quantum theory of wormholes.  Phys. Letts {\bf {106B}}
77-80 (1981)
\item   P Hajicek, Quantum wormholes (I).  Choice of the classical
solutions.  Nuc. Phys. {\bf {B185}} 254-268 (1981)
\item   P Hajicek, Duality in Klein-Kaluza Theories.  BUTP-9/82
\item   P Hajicek, Exact Models of Charged Black Holes.  I:  Geometry of
totally geodesic null hypersurface.  Commun. M. Phys. {\bf {34}} 37-52 (1973)
\item  P Hajicek, Exact Models of Charged Black Holes II:  Axisymmetries
Sationary Horizons.  Commun. M. Phys {\bf {34}} 53-76 (1973)
\item  P Hajicek, Can outside fields destroy black holes.  J. Math. Phys.
{\bf {15}} 1554 (1974)
\item  D Harari and C Lousto, Repulsive gravitaional effects of global
monopoles. Buenos Aires preprint GTCRG-90-4
\item  W A Hiscock, Magnetic Monopoles and evaporating black holes.
Phys. Rev. Letts {\bf {50}} 1734-1737 (1982)
\item  W A Hiscock, On black holes in magnetic universes.  J.
Math. Phys. {\bf {22}} 1828-1833 (1981)
\item  W A Hiscock, Magnetic monopoles and evaporating black holes.
Phys. Rev. Lett {\bf {50}} 1734-1737 (1983)
\item  W A Hiscock, Astrophysical bounds on global monopoles.
\item  W A Hiscock, Gravitational particle production in the formation
of global monopoles.
\item  P A Horvathy, Bogomolny-type equations in curved space.  Proc.
2nd Hungarian Relativity Workshop (Budapest 1987)  ed. Z Peres, World
Scientific
\item H S Hu, Non existence theorems for Yang-Mills fields and harmonic
maps in the Schwarzschild spacetime (I).  Lett. Math. Phys. {\bf {14}} 253-262
(1987)
\item H S Hu and S Y Wu, Non existence theorems for Yang-Mills fields
and harmonic maps in the Schwarzschild spacetime (II).  Lett Math. Phys. {\bf
{14}} 343-351 (1987)
\item W Israel and K A Khan Collinear paricles and Bondi dipoles in general
relativy. Nuovo Cimento {\bf33} 331 (1964)
\item   W Kinnersley and M Walker, Uniformly accelerating charged mass in
general relativity.  Phys. Rev. {\bf {D2}} 1359-1370 (1970)
\item L M Krauss and F Wilczek, Discrete gauge symmetry in continuum theories.
Phys Rev Letts {\bf 62} 1221-1223 (1989)
\item  H P Kunzle and A K M Masood-ul-Alam, Spherically symmetric static
$SU(2)$ Einstein-Yang-Mills fields.
J. Math. Phys. {\bf {31}} 928-935 (1990)
\item  A S Lapedes and M J Perry, Type D Gravitational instantons.  Phys.
Rev. {\bf {D24}} 1478-1483 (1983)
\item  D Lohiya, Stability of Einstein-Yang-Mills Monopoles and Dyons.
Ann. Phys. {\bf {14}} 104-115 (1982)
\item  M Magg, Simple proof for Yang-Mills instability.  Phys. Letts {\bf
{74B}} 246-248 (1978)
\item  E Malec and P Koc, Trapped surfaces in monopole-like Cauchy data
of Einstein-Yang-Mills-Higgs equations.  J. Math. Phys. {\bf {31}} 1791-1795
(1990)
\item   J E Mandula, Classical Yang-Mills potentials. Phys. Rev. {\bf
{D14}} 3497-3507 (1976)
\item   J E Mandula, Color screening by a Yang-Mills instability.  Phys.
Letts {\bf {67B}} 175-178 (1977)
\item  J E Mandula, Total Charge Screening.  Phys. Lett. {\bf {69B}}
495-498 (1977)
\item  E A Martinez and J W York, Thermodynamics of black holes and
cosmic strings.  IFP-342 UNC:  May 1990
\item  A K M Masood-ul-Alam and Pan Yanglian (Y L Pan)  Non Existence
theorems for Yang-Mills fields outside a black hole of the Schwarschild
spacetime.  Lett in Math. Phys. {\bf {17}} 129-139 (1989)
\item  M A Melvin, Pure magnetic and electric geons.  Phys. Lett {\bf
{8}} 65-67 (1964)
\item  C W Misner, Harmonic maps as models for physical theories.
Phys. Rev {\bf D18} 4510-4524 (1978)
\item  M J Perry, Black holes are coloured.  Phys. Letts {\bf {71B}} 234
(1977)
\item J Preskill and L M Krauss, Local discrete symmetry and quantum-mechanical
hair.
Nucl. Phys. {\bf B341}50-100 (1990)
\item J Preskill Quantum Hair. Caltech preprint CALT-68-1671 (1990)
\item  D Ray, Solutions of coupled Einstein-$SO(3)$ gauge field
equations.  Phys. Rev. {\bf {D18}} 1329-1331 (1978)
\item  P Ruback A New Uniqueness Theorem for Charged Black Holes.
Class. and Quant. Grav. {\bf{5}} L155-L159 (1988)
\item  R F Sawyer, The possibility of a static scalar field in the
Schwarzschild geometry.  Phys. Rev. {\bf {D15}}, 1427-1434 (1977).  Erratum Phys.
Rev. {\bf {D16}} (1977) 1979
\item  P Sikivie and N Weiss, Screening Solutions to Classical
Yang-Mills theory.  Phys. Rev. Letts {\bf {40}} 1411-1413 (1978)
\item  N Straumann and Z-H Zhou, Instability of colored black hole
solution.  Phys. Letts {\bf {141B}} 33-35 (1990)
\item  N Straumann and S-H Shou, Instability of the Bartnik-McKinnon
solutions of the Einstein-Yang-Mills equations.  Phys. Letts {\bf {237}} 353-356
(1990)
\item  P van Nieuwenhuizen, D Wilkinson and M J Perry, Regular solution
of 't Hooft's magnetic monopole in curved space.  Phys. Rev. {\bf {D13}} 778-784
(1976)
\item  M Y Wang, A solution of coupled Einstein-SO(3) gauge field
equations. Phys. Rev. {\bf {D12}} 3069-3071 (1975)
\item  M S Volkov and D V Gal'tsov, Non-Abelian Einstein-Yang-Mills
black holes.  JETP.  Letts {\bf {50}} 346-350 (1989)
\item  D V Galt'sov and A A Ershov.  Yad.  Fiz.  {\bf {47}} 560 (1988)

\vfill
\eject

\bye